\documentclass[12pt]{article}

\usepackage{fullpage}
\usepackage{setspace}
\usepackage{graphicx}
\usepackage{epsfig} 
\usepackage{epstopdf}

\usepackage[usenames,dvipsnames]{color}
\usepackage[usenames,dvipsnames,svgnames,table]{xcolor}
\usepackage[normalem]{ulem} 
\usepackage{amsmath}  
\usepackage{latexsym}
\usepackage{multirow} 
\usepackage[super,comma,sort&compress]{natbib}

\usepackage{algorithm}
\usepackage[noend]{algpseudocode}
\usepackage{amsfonts}

\usepackage{arydshln} 
\DeclareMathAlphabet{\mathpzc}{OT1}{pzc}{m}{it}

\definecolor{gray}{gray}{0.8}
\definecolor{darkgray}{gray}{0.6}

           \newcommand{\Hb}{{\mathbf H}}
           \newcommand{\Ib}{{\mathbf I}}

           \newcommand{\Mb}{{\mathbf M}}

           \newcommand{\Tb}{{\mathbf T}}
\newcommand{\ub}{{\mathbf u}}           
\newcommand{\vb}{{\mathbf v}}           
\newcommand{\wb}{{\mathbf w}}

\usepackage{dsfont}

\DeclareMathAlphabet{\mathpzc}{OT1}{pzc}{m}{it}

\newcommand{\tr    }{{\top}}

\newcommand{\ASU  }{{\mbox{\small ASU}}}

\title{Use Symmetry to Elucidate the Roles of Global Shape and Local Interactions in Protein Dynamics and Cooperativity} 

\author{
Guang Song$^{1,2*}$ \\
  \\
  $^1$Department of Computer Science, \\
  $^2$Program of Bioinformatics and Computational Biology, \\
  Iowa State University, Ames, IA 50011, USA \\
  \\
  $^*$Correspondence to: Guang Song, Tel: 515-294-1696; \\
  Fax: 515-294-0258; E-mail: gsong@iastate.edu
}

\begin{document}
\maketitle

\begin{abstract}
Shape had been intuitively recognized to play a dominant role in determining the global motion patterns of bio-molecular assemblies. However, it is not clear exactly how shape determines the motion patterns. What about the local interactions that hold a structure together to a certain shape? The contributions of global shape and local interactions usually mix together and are difficult to tease part.
In this work, we use symmetry to elucidate the distinct roles of global shape and local interactions in protein dynamics.  
Symmetric complexes provide an ideal platform for this task since in them the effects of local interactions and global shape are separable, 
allowing their distinct roles to be identified.
Our key findings based on symmetric assemblies are:
(i) the motion patterns of each subunit are determined primarily by intra-subunit interactions (IRSi), and secondarily by inter-subunit interactions (IESi); 
(ii) the motion patterns of the whole assembly are fully dictated by the global symmetry/shape and have nothing to do with local iESi or IRSi. This is followed by a discussion on how the findings may be generalized to complexes in any shape, with or without symmetry.

\end{abstract}

Keywords: symmetry; shape; degeneracy; cooperativity;  local interactions; phase shift; vibrations; network models

\section{Introduction}


Understanding the mechanism by which a molecular system functions is not only of great scientific interests but also 
has significant practical implications such as in drug design.  For example, in the development of HIV-1 anti-viral drugs, 
modern drug design has become largely mechanistic-based~\cite{RN968}. Structural and mechanistic understandings gained from basic research have stimulated much hope in drug development~\cite{RN966}. 

The technological advances in structure determination and the increasing availability of atomic or near-atomic structures of large molecular systems as a result have been instrumental in our understanding of the functional mechanisms of bio-molecular systems. Since bio-molecular systems are constantly in motion, another important piece of information that is often needed in uncovering the mechanism is dynamics. 
Combined together, structure and dynamics provide a fuller
understanding of the mechanism. 
It is intuitively known that there is an intrinsic relationship among structure, dynamics, and function. 
However, the relationship is not well understood.  
It is not fully clear why a given molecular system
is structured in a certain way and what dictates its dynamics or motion patterns.

Our objective in the work is to 
elucidate what determines the global motion patterns of a biomolecular assembly at around its equilibrium state. 
We will employ normal mode analysis (NMA)~\cite{brooks83,levitt83,go83} as the approach, since 
NMA is best suited for studying the equilibrium dynamics and global motion patterns, and for understanding the relationship between structure and dynamics.

Regarding the question what determines the global motion patterns of a bio-molecular system, 
shape had been intuitively recognized to play a dominant role~\cite{Ming02a,Lu05role,Tama06symmetry,FredSchulten08,Kurkcuoglu08}. However, it is not fully clear exactly how shape influences the motion patterns. For one thing, even shape itself is not well defined. For another, what about the local interactions that define and hold the structure together to a certain shape? What is their contribution or what difference do they make to the dynamics?

Much of our understanding that shape determines the overall
dynamic patterns was based on empirical evidence, as observed by Tama and Brooks, who noted that ``The evolution of the techniques and models
for NMA, from an all-atom representation
of the biological molecule to multi-resolution
EN models, has shed light on and new insights
into the role of shape and form in controlling
the motions of biological molecules
and their assemblies.''~\cite{Tama06symmetry}. 
In another work, the authors applied randomization to the Hessian matrix elements while maintaining the structure of the matrix (i.e., keeping the zero elements zero and non-zero elements non-zero).
The low-frequency normal modes were found to be robust to the 
randomization. They then concluded that the global shape of molecules played a more dominant role in determining molecular motions than
local interactions. 
However, the approach used only two proteins and exploited randomization only within a small range of spread of the parameters~\cite{Lu05role}. 

In this work,
we aim to characterize precisely the motion patterns of an 
assembly as a whole and the motion patterns of subunits as components, and exactly what are the underlying determinants
of these motion patterns. 
What is the role of shape? 
What is the role of local interactions, specifically inter-subunit interactions (IESi) or intra-subunit interactions (IRSi),
and how do they affect the motions of each subunit individually and the assembly as a whole? 
Lastly, we address the issue of cooperativity, i.e., to what extent do the subunits correlate or synchronize their motions? How much does cooperativity among the subunits depend on the strength of inter-subunit interactions?

To answer these questions, we first turn to assemblies with symmetry, for the following reasons. Though it is generally accepted that shape determines the motion patterns of single-chain structures or 
structure assemblies~\cite{Ming02a,Lu05role, Tama06symmetry, FredSchulten08,Kurkcuoglu08}, unlike volume or weight, shape 
is generally not well defined geometrically~\cite{Liang98a,Liang98b}.
According to wikipedia (\url{https://en.wikipedia.org/wiki/Shape}), ``A shape is the form of an object or its external boundary, outline, or external surface, as opposed to other properties such as color, texture or material composition.'' The definition does not exactly quantify what a shape it. 

Fortunately, there are some special cases where shape is precisely defined. One such case is when shape
is associated with symmetry. To say an assembly has a symmetric shape or it has a certain point 
group symmetry means the same thing. One may thus say a certain symmetry rigorously defines 
a certain shape or a certain attribute of the shape. For example, the cyclic symmetry defines a ring shape. the dihedral symmetry define a double-ring shape. Symmetry can be thought of as a special kind of ``shape'' that has a precise definition. 

It is helpful to note that shape in biological systems is often associated with symmetry.  In protein data bank (PDB)~\cite{berman00}, nearly 40\% of the entries are complexes with global symmetry. Symmetry is useful structurally in constructing complexes with certain structural 
shapes, such as rings, double-rings, or spherical containers that are often used to store viral genomes or build bacterial micro-compartments~\cite{Klose16,Sutter17}, etc.  Symmetry allows large complexes to be assembled out of a small number of protein types. 
Symmetry is important functionally as some functional processes rely heavily on the symmetry. Take the AAA+ family for example, most of the proteins in this family are ring-shaped and have symmetric motions that are used to thread proteins through their central pore~\cite{Hanson05,Na16c}. 
Computationally, symmetry is useful as it can be taken advantage of to significantly accelerate the normal mode computations~\cite{Simonson92,Vlijmen01,Vlijmen05,Kidera12,Na16c, Song17a,Song17b, Song17c, Song18a}. 
Moreover, as will be shown next,
symmetric complexes provide an ideal platform to elucidate the roles of global shape and local interactions. 
These roles mix together in non-symmetric systems and are difficult to tease apart. 
In symmetric complexes,  the effects of local interactions and global shape become separable, 
allowing their roles to be clearly identified. 

\section{Methods and Results}


\subsection{The distinct roles of global shape and local interactions in the motions of symmetric complexes}
In this section, we will show that the global shape and local interactions have distinct roles in determining the motions of symmetric complexes. 

Given a symmetric complex of any symmetry group that has $h$ asymmetric subunits (or $\ASU$s), its Hessian matrix in each $\ASU$'s local coordinates~\cite{Song17a,Song17b} 
is given as follows:
\begin{equation}
\Hb^{local} = 
\begin{bmatrix}
\Hb_{0}     & \Hb_{1} & \dots  & \Hb_{h-2} & \Hb_{h-1}  \\
\vdots  & \vdots &	\ddots    & \vdots  & \vdots   \\
\end{bmatrix},
\label{eq:any}
\end{equation}
where $\Hb_{0}$ represents intra-subunit interactions and $\Hb_{j}$ inter-subunit interactions between the current subunit (or the $0^{th}$ subunit) and the $j^{th}$ subunit.  Note that due to symmetry, the block elements on the other rows of $\Hb^{local}$ (not listed) are the same as those on the first row except the order is different. {\em This is true for all symmetry groups.} 

What also is true for all symmetry groups is that there exists a symmetry transformation $\Tb$,  under which the Hessian matrix $\Hb^{local}$ becomes block diagonal~\cite{Cotton90}.
That is, 
\begin{equation}
\Hb^{sym} = \Tb^{\tr}\Hb^{local} \Tb,
\label{eq:sym}
\end{equation}
where $\Hb^{sym}$ is the Hessian matrix in symmetry coordinates~\cite{Cotton90} and is block diagonal. $\tr$ denotes transpose.  $\Tb$ is unique for each symmetry group and is solely determined by representation matrices of the symmetry group. In the following, we will show how this works out particularly for assemblies with cyclic symmetry or icosahedral symmetry. 

\paragraph{Assemblies with the cyclic symmetry.} 
Consider a ring-like assembly that has $h$ identical subunits and cyclic symmetry. Its Hessian matrix in local coordinates is a circulant matrix~\cite{Song17a}:
\begin{equation}
\Hb^{local} = 
\begin{bmatrix}
\Hb_{0}     & \Hb_{1} & \dots  & \Hb_{h-2} & \Hb_{h-1}  \\
\Hb_{h-1}     & \Hb_{0} & \dots  & \Hb_{h-3} & \Hb_{h-2}  \\
\vdots  & \vdots &	\ddots    & \vdots  & \vdots   \\
\Hb_{2}     & \Hb_{3} & \dots  & \Hb_{0} & \Hb_{1}  \\
\Hb_{1}     & \Hb_{2} & \dots  & \Hb_{h} & \Hb_{0}  \\
\end{bmatrix},
\label{eq:Hc}
\end{equation}
where $\Hb_{0}$ represents intra-subunit interactions and $\Hb_{j}$ inter-subunit interactions between the current subunit and the $j^{th}$ subunit down the ring. 
$\Hb_{h-j} = \Hb^{\tr}_{j}$ (the transpose of $\Hb_j$) since the matrix has to be symmetric. 
 
The inter-subunit interactions (IESi), $\Hb_j$ ($1 \le j \le h-1$), generally are substantially weaker than intra-subunit interactions (IRSi), or $\Hb_0$.

Based on the character table of cyclic symmetry, one can write down a transformation matrix $\Tb$~\cite{Song17a}, 
\begin{equation}
\Tb = \frac{1}{\sqrt{h}}
\begin{bmatrix}
1    & 1 & 1 & \dots  & 1  \\
1 & \omega_h    & \omega_h^2 & \dots        & \omega_h^{h-1}  \\
1 & \omega_h^2    & \omega_h^4 & \dots        & \omega_h^{2(h-1)}  \\
\dots  & \dots & \dots    & \dots  & \dots   \\
1 & \omega_h^{h-1}    & \omega_h^{2(h-1)} & \dots        & \omega_h^{(h-1)^2}  \\
\end{bmatrix}
\label{eq:F}
\end{equation}
where $\omega_h$ is $e^{\frac{2\pi i}{h}}$ and $h$ is the number of subunits,
such that when applied to the original Hessian matrix, produce the Hessian matrix in
the symmetry coordinates that is block diagonal, i.e., 
\begin{equation}
\Hb^{sym} = \Tb^\tr \Hb^{local} \Tb = diag(\Mb_0, \Mb_1, \cdots, \Mb_{h-1}),
\label{eq:cycH}
\end{equation}
where $\Mb_j = \sum_{k=0}^{h-1} \omega_h^{kj}\Hb_k$. 

The motion pattern of a subunit is thus dictated by $\Mb_j$'s ($0 \le j \le h-1$), which can be rewritten as:
\begin{equation}
\Mb_j = \Hb_0 + \sum_{k=1}^{h-1} \omega_h^{kj}\Hb_k. 
\label{eq:mj}
\end{equation}
In the above equation, the first term $\Hb_0$ represents intra-subunit interactions (or IRSi), while the second term inter-subunit interactions (or IESi).
Since the second term consisting of non-bonded interactions is generally {\em much weaker} than the first term that is composed mostly of bonded interactions, the motion pattern of each subunit, $\Mb_j$, is thus determined by 
the intra-subunit interestions (IRSi) {\em primarily} and inter-subunit interactions (IESi) {\em secondarily}. 


Assume ${\ub_j}$ is one of the eigenvectors of $\Mb_j$, 
then 
$\wb_j$ = $(0_{3N}, 0_{3N}, \cdots, {\ub_j}^T, \cdots, 0_{3N})^T$ is an eigenvector of $\Hb^{sym}$ in eq.~\ref{eq:cycH},  
where $0_{3N}$ represents a row vector with $3N$ zeros and $N$ is the number of atoms in each subunit.
The normal mode of the whole assembly in the local coordinates is~\cite{Song17a}, 
\begin{equation}
\vb = (\Tb\otimes I_{3N})\wb_j
 = 
\begin{bmatrix}
\ub_j \\
\ub_j\omega_h^j\\
\vdots \\
\ub_j\omega_h^{j(h-1)}\\
\end{bmatrix}
\label{eq:v}
\end{equation}
Eq.~(\ref{eq:v}) is the general form of the normal modes of assemblies with cyclic symmetry. It clearly shows that different subunits have exactly the same motions except for 
a constant phase shift (or phase lag). Thus
the motion patterns of the whole assembly are standing waves, in which the motions of the subunits are the same but there is a phase shift/lag between neighboring subunits. 
It is worth noting that
this global motion pattern is dictated solely by $\Tb$ (Eq.~\ref{eq:F}), the same transformation matrix that is shared by all complexes with cyclic symmetry and $h$ subunits (or $C_h$ in short), having nothing to 
do with IESi or IRSi. 
For different complexes with the same cyclic symmetry, the difference in their subunit 
compositions affects {\em only} the the motions patterns of individual subunits, {\em not} the motion patterns of the assembly as a whole.

{\bf Summary.} 
The above results regarding complexes of cyclic symmetry can be summarized as follows:
\begin{itemize}
\item The motion patterns of each subunit are determined primarily by intra-subunit interactions (IRSi), and secondarily by inter-subunit interactions (IESi), which serve mostly as a perturbation. 
\item The motion patterns of the whole assembly are fully dictated by the symmetry. Symmetry dictates that the overall motion patterns are standing waves. Surprisingly, neither iESi nor IRSi affects the global motion patterns. 
\end{itemize}
A third issue not included above is the extent of cooperativity: how stable are the motion patterns? How strongly do subunits maintain their cooperativity with one another?  Some motion patterns may be highly vulnerable to degeneracy and may disappear at slight disturbance in structure~\cite{Na16b}. We will address the issue of cooperativity later in the text.

\paragraph{Assemblies with the icosahedral symmetry.}
For complexes with icosahedral symmetry, 
the Hessian matrix in local coordinates takes the same form as Eq.~(\ref{eq:any}) with $h$ being 60. The order of the blocks in other rows is known and can be found in Ref.~\citenum{Song18a}. 

Similar to complexes with cyclic symmetry, there exists a transformation matrix $\Tb$ (also given in Ref.~\citenum{Song18a}) under which the Hessian matrix becomes block diagonal. {$\Tb$ is solely determined by 
representation matrices of the icosahedral point group and has nothing to do with the structural composition of 
the subunits nor their interactions.  }

That is~\cite{Song18a}, 
\begin{equation}
\Hb^{sym} = \Tb^{\tr}\Hb^{local} \Tb = 
\begin{bmatrix}
\Hb_E & 0 & 0 & 0 & 0 \\
0 	& \Ib_3 \otimes \Hb_{T1} & 0 & 0 & 0 \\
0   &  0  & \Ib_3 \otimes \Hb_{T2} & 0 & 0 \\
0   &  0  &  0 &  \Ib_4 \otimes \Hb_G & 0 \\
0   &  0  &  0 & 0 & \Ib_5 \otimes \Hb_H \\
\end{bmatrix},
\label{eq:Hicosa}
\end{equation}
where $\Hb^{sym}$ is the Hessian matrix of icosahedral complexes in symmetry coordinates. The diagonal blocks correspond to the five {\em irreps} of the icosahedral point group~\cite{Song18a}.  

Again, the motions of each subunit are dictated by the diagonal blocks
$\Hb_E$, $\Hb_{T1}$, $\Hb_{T2}$, $\Hb_G$,  and $\Hb_H$ (whose subscripts correspond to the five {\em irreps} of icosahedral point groups, namely $E$, $T_1$, $T_2$, $G$, and $H$), all of which can be expressed as a summation of an intra-$\ASU$ interaction term and an inter-$\ASU$ interaction term~\cite{Song18a}. For example,
\begin{equation}
\Hb_{E} = \Hb_{0} + \sum_{j=1}^{h-1}\Hb_{j}
\end{equation}
where $\Hb_{0}$ (Eq.~\ref{eq:any}) represents interactions within the first $\ASU$ while the second term represents the interactions between the first $\ASU$ and the other $\ASU$s.  

On the other hand, the transformation matrix $\Tb$ of icosahedral symmetry, which depends solely on the symmetry and has nothing to with IESi nor IRSi, determines 
the global motion patterns 
in a similar fashion as in Eq.~(\ref{eq:v}). $\Tb$ determines how the motion patterns of individual subunits 
are to be pieced together to form the normal mode of the whole 
assembly. To put this in an analogy, if the motion patterns of a subunit were a tile and the motion patterns of the whole assembly were a floor of tiles, $\Tb$ contains the rule for laying out the tiles in order to achieve a global ``floor pattern'' that is consistent with the symmetry of the assembly.

The same conclusion found for complexes with cyclic or 
icosahedral symmetry clearly holds on complexes with other symmetries as well. As shown in Eq.~(\ref{eq:sym}), 
{for every symmetry group, there exist a set of symmetry coordinates under which the Hessian matrix, $\Hb^{sym}$, is block diagonal, and a transformation matrix $\Tb$ that relates $\Hb^{sym}$ with $\Hb^{local}$.} 
As a result, 
the local motion patterns of each subunit is determined {\em primarily} by IRSi within the subunit and {\em secondarily} by IESi, while the global motion patterns of the whole complex are pieced together from component (or subunit) motions 
according to a rule expressed in $\Tb$ that is determined solely by the global symmetry of the complex.

\subsection{Use symmetry to understand cooperativity}
The functional processes of many biological systems require cooperativity.
Understanding cooperativity is thus highly important~\cite{CuiKarplus08}.
The range of cooperativity has been studied using elastic 
network models and normal mode analysis~\cite{Yang09}. 
In this work, when speaking of cooperativity, we focus on the following specific aspect of it:
How strongly correlated are the motions of subunits in a large assembly? What are key factors that influence this cooperativity? One of our findings is that cooperativity among subunits is frequency-dependent, i.e., cooperativity among subunits is stronger at some frequencies and weaker at others. 

\paragraph{Inter-subunit interactions effect cooperativity among subunits.}
In the following, we show how the extent of cooperativity among subunits is influenced by inter-subunit interactions (IESi). 

Recall that the motions of a subunit are dictated by 
intra-subunit interactions primarily and inter-subunit interactions secondarily.
Without IESi, all subunits have the same vibrational spectrum 
and the vibrations of the whole assembly is fully degenerate, and there is 
no cooperativity among the subunits.  
 Inter-subunit interactions serve as a perturbation and cause 
 the otherwise degenerate modes of the subunits to mix together.
 As a result, the otherwise degenerate vibrational frequency lines are split and the modes become mostly non-degenerate. Apparently, the extent of split depends on the magnitude of the perturbation, or the strength of inter-subunit interactions. If the magnitude of perturbation is very small, the split will be narrow and under slight structure variations, these modes become effectively degenerate again~\cite{Na16b} and cooperativity is lost as a result. 
 
To illustrate this quantitatively, we consider the simplest assembly with two subunits: a dimer. The Hessian matrix of a symmetric dimer when written in local 
coordinates is~\cite{Song17a,Song17b}:
\begin{equation} 
\Hb^{local} = 
\begin{bmatrix}
\Hb_{0} & \Hb_{1} \\
\Hb_{1} & \Hb_{0} \\
\end{bmatrix},
\label{eq:Hlocal}
\end{equation}
where $\Hb_1$ represents inter-subunit interactions. $\Hb^{local}$ becomes 
block diagonal when transformed to symmetry coordinates~\cite{Cotton90, Song17b}:
\begin{equation} 
\Hb^{sym} = 
\begin{bmatrix}
\Hb_{0} + \Hb_{1}  & 0  \\
0 & \Hb_{0} - \Hb_{1} \\
\end{bmatrix}.
\label{eq:Hsym}
\end{equation}

Now let $\vb^{(0)}_i$ and $\lambda_i$ (where $1 \le i \le 3N$ and  $N$ is the number of atoms in each subunit) represent respectively the eigenvalues and eigenvectors of $\Hb_0$. If $\Hb_1 = 0$, i.e., if there is no inter-subunit interactions, $\Hb^{sym}$ becomes,
\begin{equation} 
\Hb^{sym} = 
\begin{bmatrix}
\Hb_{0}   & 0  \\
0 & \Hb_{0}  \\
\end{bmatrix}.
\label{eq:Hsym0}
\end{equation}
In such a case, all the normal modes of the dimer are obviously doubly-degenerate: the two subunits have identical vibrational spectra.     

Inter-subunit interactions $\Hb_1$ remove this degeneracy.  
Indeed, inter-subunit interactions $\Hb_1$ in Eq.~(\ref{eq:Hsym}) cause
the otherwise degenerate frequency lines to split. 
The extent of split 
can be well estimated by the zero-order approximation according to the
perturbation theory~\cite{Landau77, wiki-perturbation} by:
\begin{equation}
\Delta \lambda_i = {\vb_i^{(0)}}^\tr \Hb_1 {\vb_i^{(0)}}
\label{eq:dH}
\end{equation}
That is, instead of having two degenerate modes corresponding to the eigenvalue $\lambda_i$,
there are two split modes of eigenvalues $\lambda_i +  \Delta \lambda_i$ and   $\lambda_i -  \Delta \lambda_i$.

\begin{figure}[H]
\centerline{\includegraphics[width=0.8\textwidth]{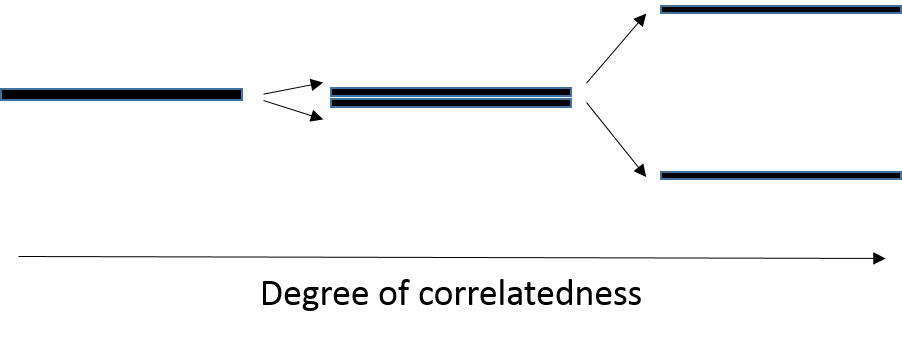}}
\caption{\small
}
\label{fig:correlation}
\end{figure}

$\lambda_i +  \Delta \lambda_i$ ($1 \le i \le 3N$) are the eigenvalues of $\Hb_0 + \Hb_1$ and correspond to fully symmetric modes in which the two subunits in the dimer have identical motions in their local coordinate frames. $\lambda_i - \Delta \lambda_i$ ($1 \le i \le 3N$), on the other hand, are the eigenvalues of $\Hb_0 - \Hb_1$ and correspond to fully anti-symmetric modes in which the two subunits in the dimer have exactly opposite motions~\cite{Song17b}. In either case, the motions of the two subunits are fully correlated. Therefore, the inter-subunit interactions that cause the splitting in the spectrum cause also the motions of the two subunits in the dimer to change 
from fully uncorrelated to fully correlated. This is illustrated in Fig.~\ref{fig:correlation}. Intuitively we know this is true: that interactions should cause the motions of the subunits to correlate. 

In reality, the transition from fully uncorrelated to fully correlated is often not black and white, but has a degree of {\em correlatedness} that depends on the 
extent of the split, which in turn depends on the magnitude of inter-subunit 
interactions. This again is in line with our intuition that weak interactions among 
subunits should incur a weak correlation and strong interactions a strong 
correlation. 
The following explains why this is the case.

Since it is evident that the extent of the split $\Delta E$, which amounts to $2\Delta\lambda$, is linearly proportional to the magnitude of inter-subunit interactions $\Hb_1$, i.e.,
\begin{equation}
\Delta E = 2\Delta \lambda_i \propto ||\Hb_1||,
\end{equation} 
weaker inter-subunit interactions $\Hb_1$ should therefore lead to a smaller split. 
When the split is small, the two split modes will have nearly identical vibrational frequencies. 
According to our previous work on the effective degeneracy of proteins~\cite{Na16b}, 
normal modes with nearly identical frequencies are effectively degenerate. The degeneracy implies that each subunit can move on its own, independent of the other subunit. A larger split on the other hand
ensures a stronger entanglement of the vibrations of the subunits and their cooperativity. {\em The width of the split is thus an indicator of the 
extent of cooperativity} (see Figure~\ref{fig:correlation}).

\begin{figure}[H]
\begin{minipage}{.48\linewidth}\centerline{\includegraphics[width=\linewidth]{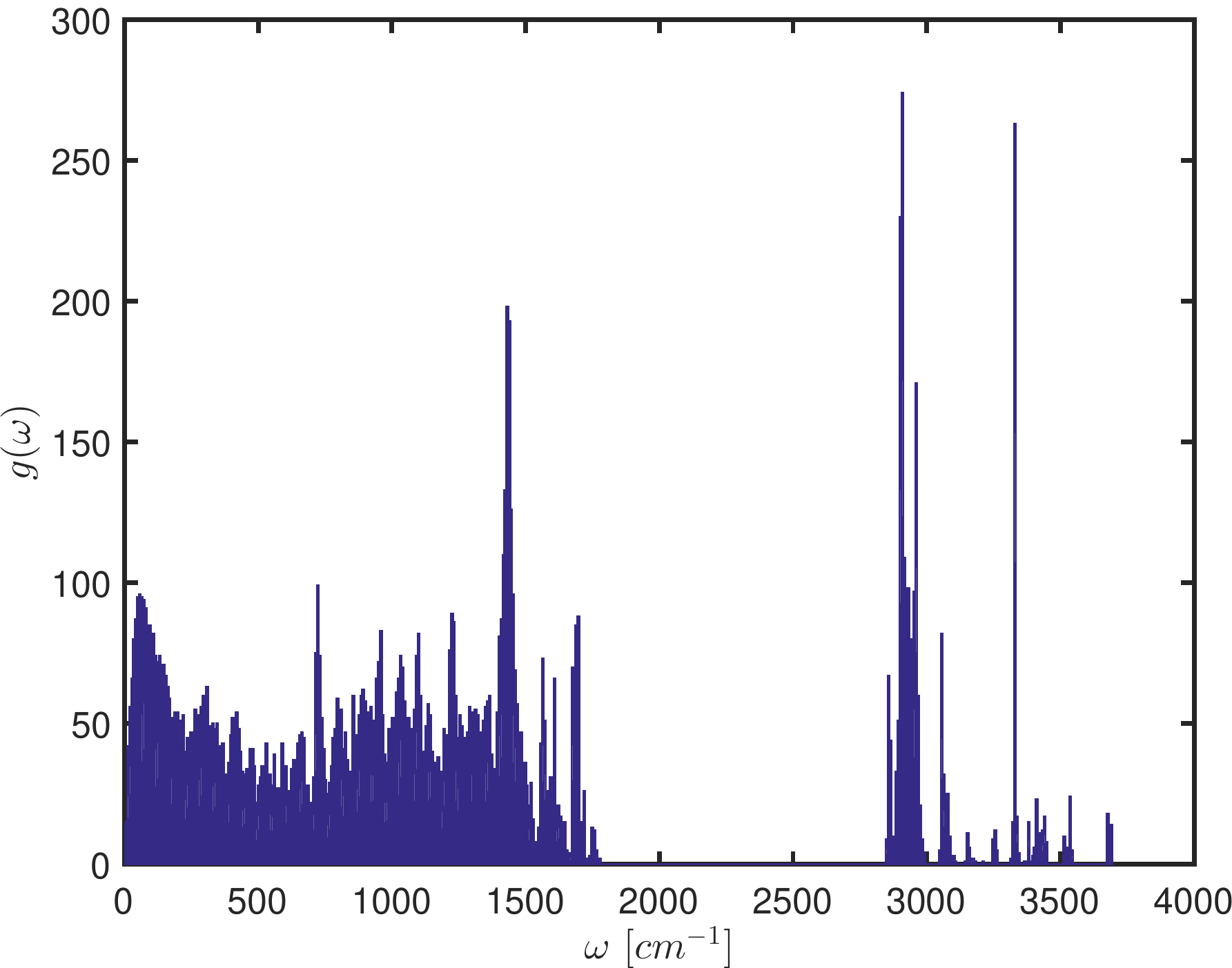}}\end{minipage}
\begin{minipage}{.48\linewidth}\centerline{\includegraphics[width=\linewidth]{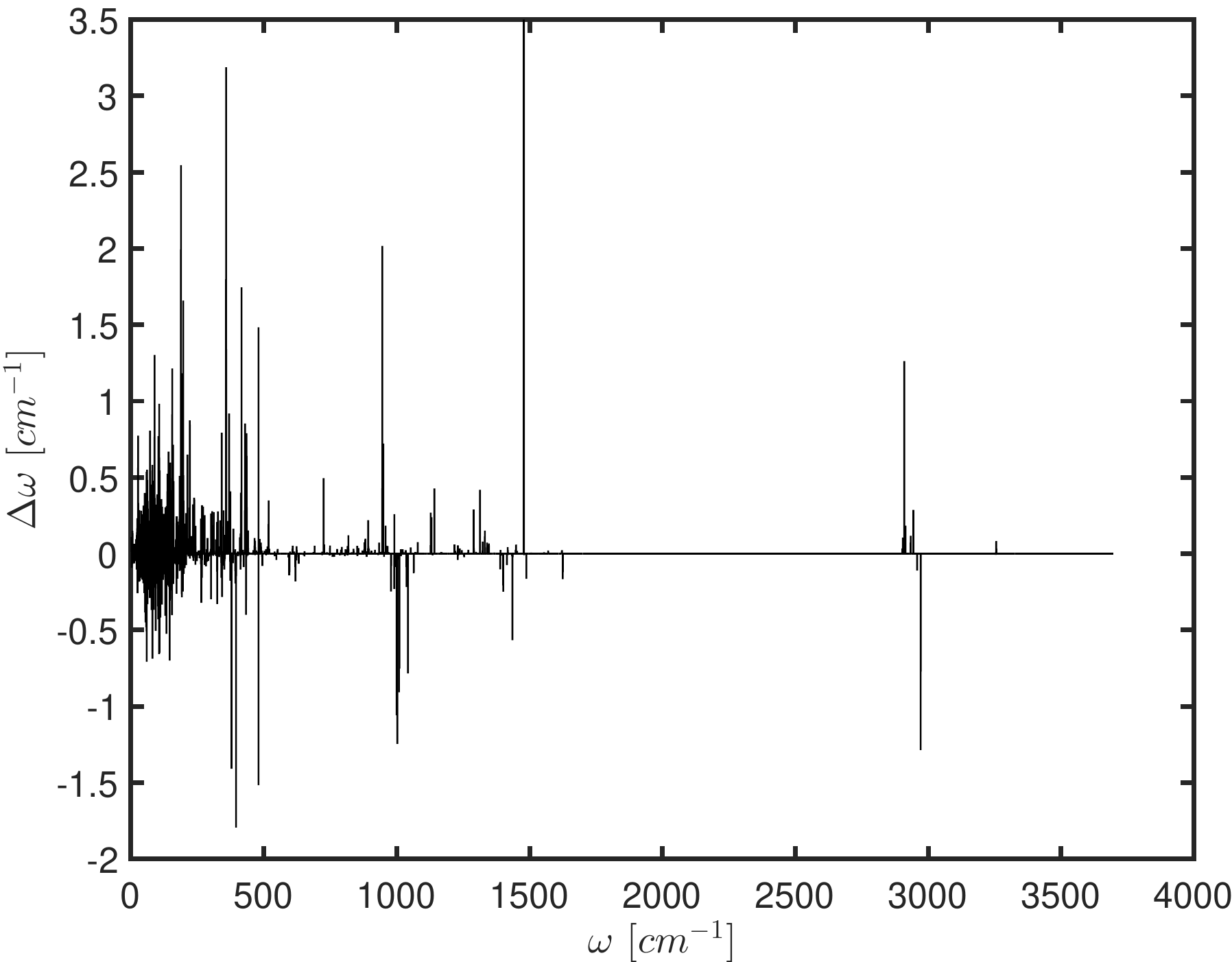}}\end{minipage}
\begin{minipage}{.48\linewidth}\centerline{(A)}\end{minipage}
\begin{minipage}{.48\linewidth}\centerline{(B)}\end{minipage}
\caption{\small
}
\label{fig:dH}
\end{figure}

\paragraph{The extent of cooperativity is frequency-dependent.} 
Figure~\ref{fig:dH} shows quantitatively how inter-subunit interactions (IESi)
influence the frequencies of the modes, using 
the crystal structure of tyrosinase from Bacillus megaterium (pdb-id: 5OAE, a dimer)~\cite{Ferro18} as an example. To this end, we apply sbNMA~\cite{Na14b} and first compute $\Delta \lambda$ using eq.~(\ref{eq:dH}). Since the split in eigenvalues is $2\Delta \lambda$ 
and the vibrational frequency $\omega = \frac{1}{2\pi c}\sqrt{\lambda}$ (in cm$^{-1}$) where $c=2.997925 \times
10^{10} cm/sec$ is the speed of light, 
the extent of split in the frequency spectrum $\Delta\omega$ is thus:
\begin{equation}
\Delta\omega_i = \frac{1}{2\pi c} \frac{\Delta\lambda_i}{\omega_i} = 
\frac{1}{2\pi c} \frac{{\vb_i^{(0)}}^\tr \Hb_1 \vb_i^{(0)}}{\omega_i}
\end{equation}

$\Delta \omega_i$ represents the zero-order correction to the frequency 
spectrum due to the existence of inter-subunits interactions. 
$\vb_i^{(0)}$ represents the eigen-modes of the systems without inter-subunit 
interactions.

On one hand,
it is clear from Eq.~(\ref{eq:dH}) that the extent of split 
is linearly proportional to the {\em magnitude} of inter-subunit interactions $\Hb_i$. 
On the other hand, the extent of split is inversely proportional to the vibrational frequency $\omega$. This frequency dependence is evident also in Fig.~\ref{fig:dH}(B), where it is seen that on average the split $\Delta\omega$ is most pronounced at the 
the low frequency range and becomes much smaller (or sparser) at high frequencies, except for a few spikes. 

The low value of $\Delta\omega$ at the high frequency end can be understood intuitively as follows. At high frequencies, most 
vibrations are highly localized, and as a result, there may be no vibration 
at the boundary between subunits. In such a case, 
inter-subunit interactions have little or no effect on the dynamics of the subunits. The extent of split is consequently very small or even none. 

Secondly, $\Hb_1$ represents contributions of the non-bonded interactions 
between subunits. These contributions were
shown to influence vibrations mostly in the frequency range of 0-500~cm$^{-1}$~\cite{Na16a}.
At this frequency range intra-subunit and inter-subunit vibrations take 
a similar frequency and consequently the subunits have the strongest correlations. That is, subunits 
resonate with one another the strongest when the inter-subunit interactions are 
able to transmit their vibrations to one another most effectively. 
Currently charge-charge interactions or hydrogen bonds are not included in the inter-subunit interactions in sbNMA~\cite{Na14b}. Their inclusion might increase the strength of IESi somewhat and consequently
the split in frequency shown in Figure~\ref{fig:correlation}.


\paragraph{Summary.}
In summary, inter-subunit interactions determine the extent of cooperativity. 
No/weak inter-subunit interactions means no/low cooperativity.
Another new finding is that the extent of cooperativity varies with frequency. 
The cooperativity should take place mostly within the  frequency range of [0, 500 cm$^{-1}$].  
Vibrations of subunits at high frequencies should be mostly uncorrelated. 
It is postulated that key function-related motion patterns should take place at the frequency range where 
cooperativity is the highest. 

\section{Discussion}
In this work, we have used symmetry to 
elucidate the roles of global shape and local interactions in protein dynamics and cooperativity. Our key findings from studying symmetric complexes are: (i) the motion patterns of each subunit are determined primarily by intra-subunit interactions (IRSi) and secondarily by inter-subunit interactions (IESi); (ii) the motion patterns of the whole assembly are fully dictated by the global symmetry and have nothing to do with iESi or IRSi.

\paragraph{From symmetry to shape.}
As aforementioned, shape in general is not well defined mathematically while symmetry is. Symmetry 
represents a special kind 
of shape (or an aspect of shape) that has a precise mathematical definition. 
A key result from the current study is that the overall motion patterns of symmetric assemblies are dictated solely by global symmetry, or global shape, 
while local symmetry or interactions influence only the motions patterns of individual subunits.
Can we generalize this 
result about complexes with symmetry to 
complexes without symmetry? Though symmetry is most often represented by a character table in group theory while a general shape cannot, fortunately both symmetry and shape can be represented by a network model. 

\begin{figure}[H]
\begin{minipage}{.48\linewidth}\centerline{\includegraphics[width=\linewidth]{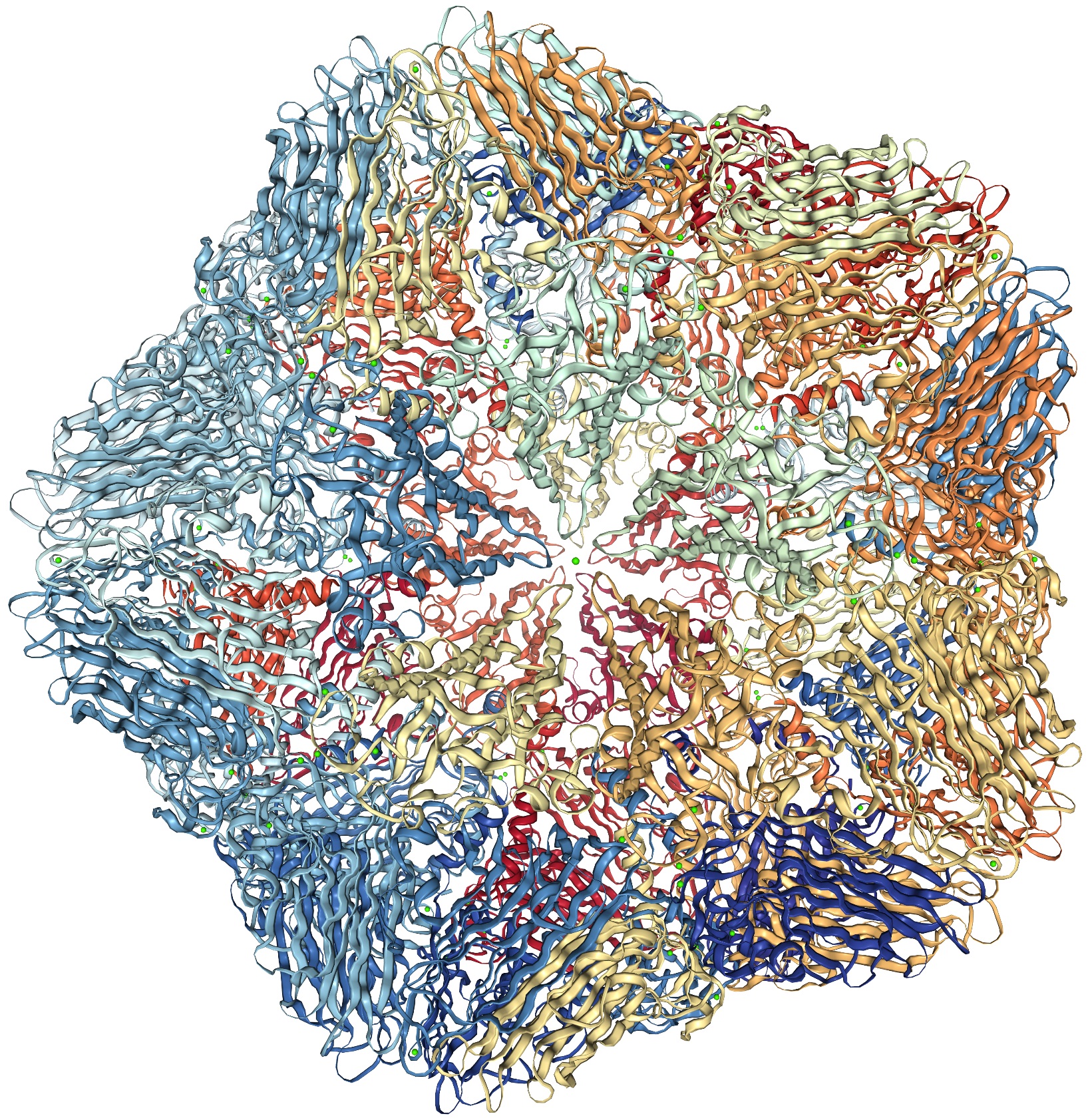}}\end{minipage}
\begin{minipage}{.48\linewidth}\centerline{\includegraphics[width=\linewidth]{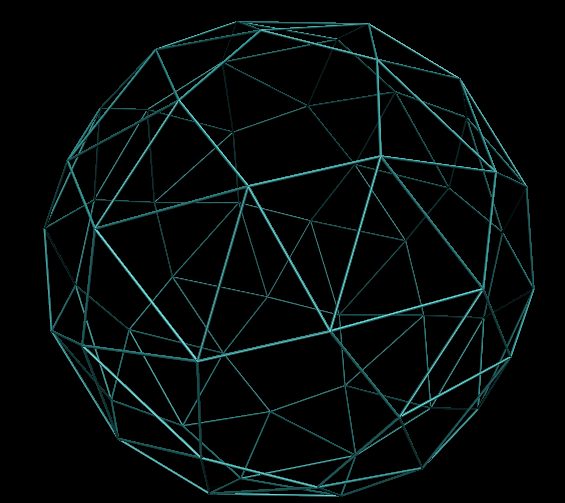}}\end{minipage}
\begin{minipage}{.48\linewidth}\centerline{(A)}\end{minipage}
\begin{minipage}{.48\linewidth}\centerline{(B)}\end{minipage}
\caption{\small
}
\label{fig:icosa}
\end{figure}
\begin{figure}[H]
\begin{minipage}{.32\linewidth}\centerline{\includegraphics[width=\linewidth]{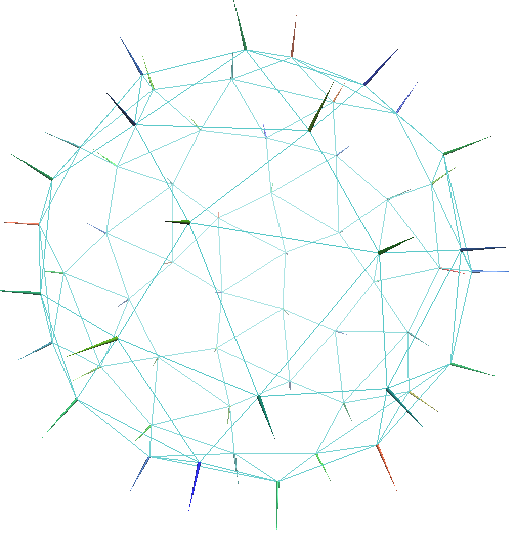}}\end{minipage}
\begin{minipage}{.32\linewidth}\centerline{\includegraphics[width=\linewidth]{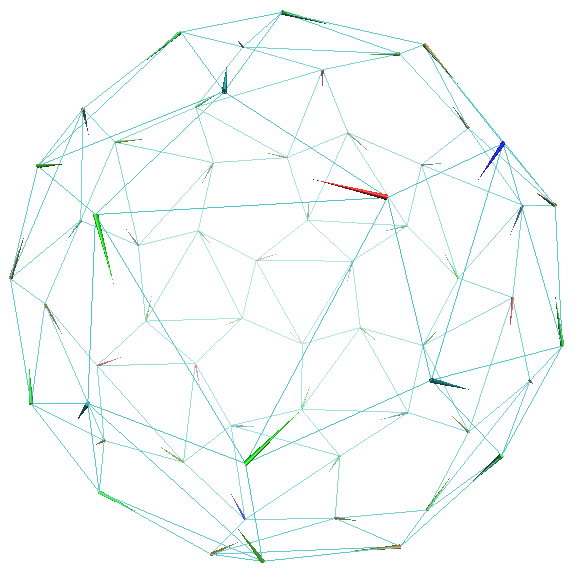}}\end{minipage}
\begin{minipage}{.32\linewidth}\centerline{\includegraphics[width=\linewidth]{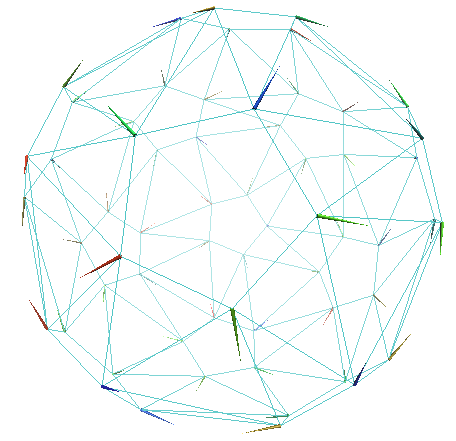}}\end{minipage}
\begin{minipage}{.32\linewidth}\centerline{(A)}\end{minipage}
\begin{minipage}{.32\linewidth}\centerline{(B)}\end{minipage}
\begin{minipage}{.32\linewidth}\centerline{(C)}\end{minipage}
\caption{\small
}
\label{fig:motion}
\end{figure}

\paragraph{Represent symmetry using a network of nodes instead 
of the character table.}
Symmetry in structures is usually 
represented by a character table and representation matrices~\cite{Cotton90,Song17b,Song18a}. However, the same symmetry, 
for instance the icosahedral symmetry possessed by a viral capsid shown in Figure~\ref{fig:icosa}(A), can 
be fully captured also by a simple network, as shown in Figure~\ref{fig:icosa}(B) where each node represents
the mass center of a corresponding protein chain of the capsid shown in Figure~\ref{fig:icosa}(A). There are 60 nodes in Figure~\ref{fig:icosa}(B) and their connectivity 
represents a shape. If we are to use this network and solve for its normal 
modes, we will obtain the same motion patterns as we get using the character table and the representation matrices (Eq.~\ref{eq:Hicosa}). That is to say, the network in Figure~\ref{fig:icosa}(B) is an alternative way to the group theory in capturing precisely the overall symmetry and consequently motion patterns of the assembly. Figure~\ref{fig:motion} shows the motion patterns of three modes (out of 180 modes) based on the network model in Figure~\ref{fig:icosa}(B). All three modes, and only these three, are singly-degenerate and fully symmetric, as is expected from the group theory that 1/60 of the modes must be
singly-degenerate and fully symmetric~\cite{Song18a}.
Now what is advantageous about this network representation (Figure~\ref{fig:icosa}(B)) is 
that it can be easily generalized to other shapes with less or no symmetry. 
This is what has been done intuitively already, for example through elastic network models~\cite{Tirion96,RN9,bahar97,Hinsen98}. 



\begin{figure}[H]
\begin{minipage}{.48\linewidth}\centerline{\includegraphics[width=\linewidth]{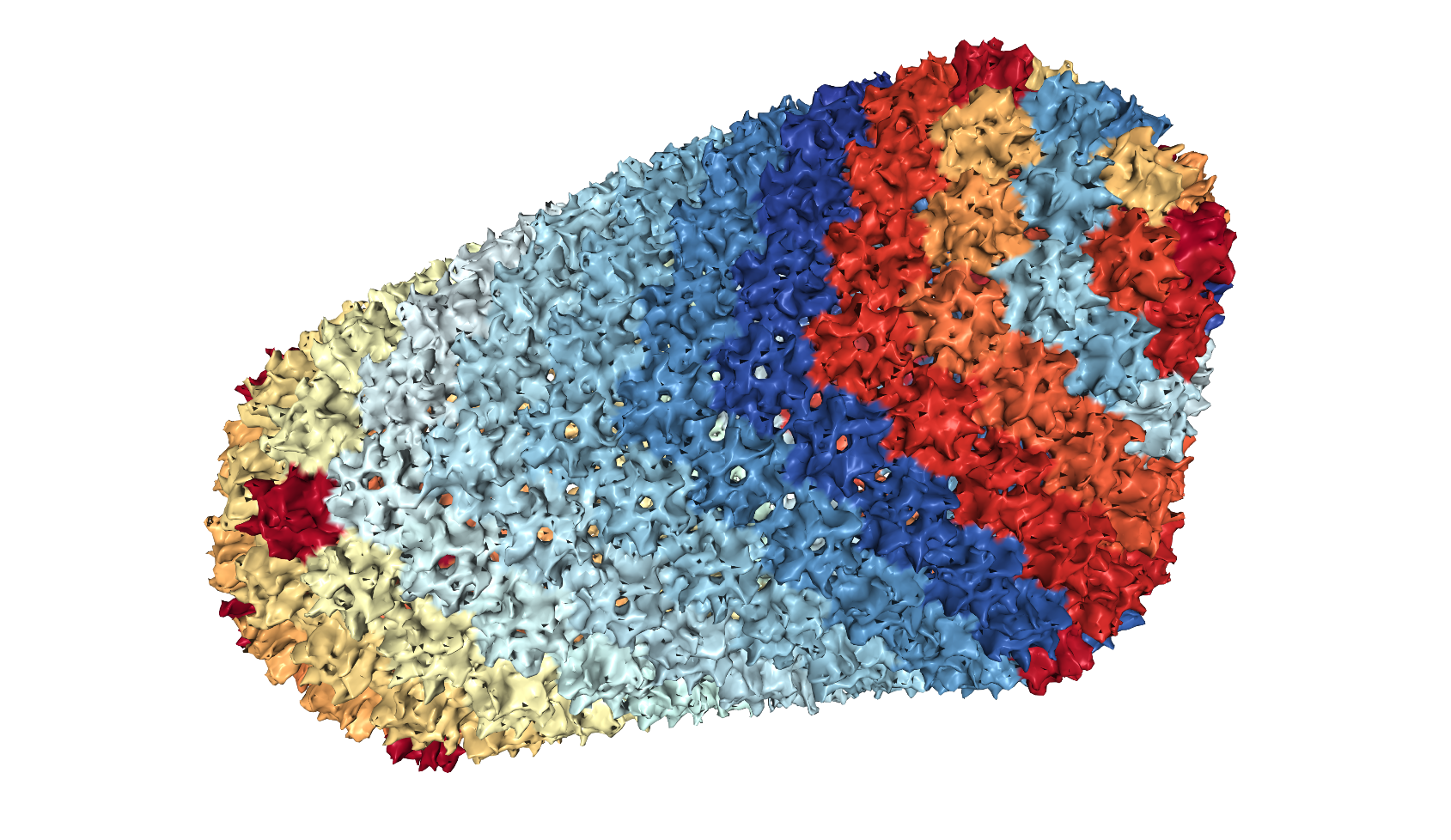}}\end{minipage}
\begin{minipage}{.48\linewidth}\centerline{\includegraphics[width=\linewidth]{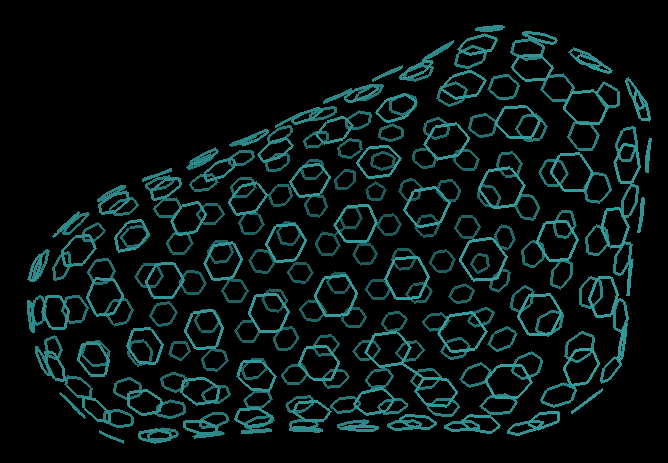}}\end{minipage}
\begin{minipage}{.48\linewidth}\centerline{(A)}\end{minipage}
\begin{minipage}{.48\linewidth}\centerline{(B)}\end{minipage}
\caption{\small
}
\label{fig:hiv}
\end{figure}

Figure~\ref{fig:hiv} shows the atomic structure of HIV-1 capsid on the left and 
its network representation on the right, with nodes placed at the mass centers of 
the 1,356 capsid proteins. By extrapolation it is reasonable to conjecture  
that just as the network in Figure~\ref{fig:icosa}(B) fully captures the global motion 
patterns of the capsid in Figure~\ref{fig:icosa} (A),
the global motion patterns of the HIV-1
capsid in Figure~\ref{fig:hiv}(A) should be captured by the network model given 
in Figure~\ref{fig:hiv}(B). 
Since the shape of HIV-1 capsid represented by the network 
of nodes in Figure~\ref{fig:hiv}(B) is ellipsoidal, the vibration of 
the whole capsid should be similar to a standing wave on an 
ellipsoidal membrane. Movies of some of these modes are given in 
Supplemental Information. Indeed, these modes show motion patterns that resemble standing waves on a membrane. 

\paragraph{The roles of shape and local interactions in determining the local or global motion patterns of a structural assembly.}
Consider the motion pattern of a large assembly that vibrates at frequency $\omega$, 
all its subunits are vibrating at the same frequency $\omega$. As shown early, the vibration pattern of each subunit is determined primarily by the IRSi (intra-subunit interactions) and secondarily by IESi (inter-subunit interactions). Since the vibration is primarily determined by IRSi, it is expected to resemble closely to its vibration at frequency $\omega$ in isolation (i.e., without IESi). This is precisely the physical basis of resonance that was observed in an earlier work~\cite{RN904,RN901}. Note that the motions of individual subunits represent only  local motion patterns, determined by IRSi primarily and IESi secondarily. 

On the other hand,
the global motion patterns of a whole assembly, which are pieced together by the motions of individual subunits,  are determined solely by global symmetry for symmetric complexes. This is most evident in complexes with cyclic symmetry, where the net effect of global symmetry is that each subunit is assigned a different phase shift (see Eq.~\ref{eq:v} and the related text). The relative phase shift among the subunits essentially captures the global motion patterns. This observation is possibly extendable to any shape in general, i.e., the global symmetry or shape of an assembly causes a phase shift to be assigned to each subunit and the relative phase shift among the subunits determines the global motion patterns. 

Besides the local motion patterns of individual subunits and the relative phase shifts assigned to them, another aspect of the global motion patterns is the relative magnitudes of motions of the subunits. The magnitude of fluctuations was shown to be inversely proportional to the packing density~\cite{Halle02}. In our terms, the relative magnitude of motions of a subunit should be roughly proportional to the magnitude of its IESi: the weaker the IESi, the larger the magnitude. IESi determines also the extent of the cooperativity between neighboring capsomeres.

In summary, for subunit $i$ that is vibrating at frequency $\omega$ within an assembly and whose vibration $\vb_i$ is denoted as 
\begin{equation}
\vb_i = \ub_i cos(\omega t + \phi_i),
\label{eq:pattern}
\end{equation}
it is reasonable to conjecture that the influence of local interactions and global symmetry or shape should come into play as follows: i)
$\ub_i$, which represents the normal mode (or the direction of motion) of the $i^{th}$ subunit as in Eq.~\ref{eq:v}, is determined primarily by IRSi and secondarily by IESi; ii) the phase shift $\phi_i$ is determined by global symmetry or shape; iii) the magnitude of $\ub_i$ is determined mostly by IESi. This conjecture is known to be true for symmetric assemblies based on the results shown earlier, but remains to be tested for other assemblies. 



\paragraph{Elastic network models.}
Insights gained in this work regarding the roles of shape and local interactions and where they come into play in a subunit's vibration expressed in Eq.~(\ref{eq:pattern}) can help NMA developers or users develop or select the right models when studying protein dynamics. 
For example, if one cares to know only the global motion patterns of a large assembly, Eq.~(\ref{eq:pattern}) indicates that a simplified network that 
captures its global symmetry or shape should be sufficient. This is the underlying basis for the success of coarse-grained elastic network models. It also reveals the models' weakness: the local motion patterns $\ub_i$ (Eq.~\ref{eq:pattern}) are often inaccurate. Their frequency spectra (or $\omega$ in Eq.~\ref{eq:pattern}) are inaccurate either, as they were found to deviate significantly from the universal spectrum of globular proteins~\cite{Na16a,dba93,Hinsen00-ms}.
On the other hand, if accurate local motion patterns are desired, e.g., when studying how the pore of a capsomere is regulated, all-atom ENM models and all-atom force fields would be necessary~\cite{Na14b, RN904, RN901}, since local dynamics 
are mostly determined by local interactions.
Mixed-grained models~\cite{Kurkcuoglu05,Kurkcuoglu09b} also seem plausible but the challenge is that an accurate force field for mixed-grained models is hard to find.

Elastic network models were developed as entropy-based models and their success was often ascribed to a good modeling of the entropy~\cite{Yang09}.
Here we see that the success of network models originates also from their accurate representation of the shape of the structures being studied, which, as the major 
determinant of the global motion patterns as is shown in this work, leads to a fairly accurate reproduction of global dynamics, as was observed also by other authors~\cite{Ming02a,Lu05role,Tama06symmetry,FredSchulten08,Kurkcuoglu08}.  

\section*{Supplemental Information}

{\bf hivExpand.mpg, hivRotate.mpg, hivContract.mpg}: movies of the normal modes of HIV-1 capsid. The modes are computed through an extremely simplified network model that represents each capsid protein as a node (thus there are 1,356 nodes in the network model). The movies show three typical motions: the radial expansion of the capsid, the rotations of capsomeres, and the contraction and expansion of the capsomeres. These motions are similar to standing waves on an ellipsoidal membrane. One of them resembles a transverse wave, while the other two longitudinal waves. 


\newpage
\section*{Figure Legends}

\subsection*{Figure 1.}
The split in spectrum lines caused by inter-subunit interactions in turn causes the motions of the subunits in a dimer to change from fully uncorrelated to fully correlated. The degree of correlatedness depends on the width of the split.  

\subsection*{Figure 2.}
{\bf (A)} The vibrational spectrum of a dimer (the crystal structure of tyrosinase from Bacillus megaterium (pdb-id: 5OAE~\cite{Ferro18}) without taking into account the effect of inter-subunit interactions, and {\bf (B)} the extent of split $\Delta \omega$ at different frequencies. 

\subsection*{Figure 3.}
The structure of STNV viral capsid (pdb-id: 4V4M~\cite{Lane11}) in {\bf (A)} atomic details (image taken from PDB~\cite{berman00}), and {\bf (B)} the same 
structure in a much simplified network model.

\subsection*{Figure 4.}
Porcupine plots of a few selected global motion patterns of STNV viral capsid as revealed by the simple
60-node network model shown in Figure~\ref{fig:icosa}{(B)}: 
{\bf (A)} radial expansion of the whole capsid, {\bf (B)} rotation along the 5-fold axes, and 
{\bf (C)} expansion or contraction of each pentamer. 
These motion patterns are in full agreement with the results that one would 
get using the group theory.

\subsection*{Figure 5.}
The structure of HIV-1 capsid in {\bf (A)} atomic details (image taken from PDB~\cite{berman00}), and {\bf (B)} the same structure in a network model, whose nodes are placed at the mass centers of 
the 1,356 capsid proteins. For clarity, only edges within each hexamer or pentamer are shown.

\newpage

\newpage
\newpage
\newpage

\newpage

\end{document}